\documentclass{appolb}

\usepackage{graphicx}
\usepackage{cite}
\usepackage{amsmath}
\usepackage{hyperref}

\newcommand{\slashed}[1]{\displaystyle{\not} #1}    

\begin{document}

\eqsec  

\title{ EFFECTS OF TOP-QUARK DECAY MODELING \\ IN $t\bar{t}\gamma$ PRODUCTION AT THE LHC
\thanks{Presented at the XXVI Cracow EPIPHANY Conference \-- \textit{LHC Physics: Standard Model and Beyond}, Cracow (Poland), 7-10 January 2020.}
}
\author{ \vspace{0.5cm} Giuseppe Bevilacqua
\address{MTA-DE Particle Physics Research Group, University of Debrecen, H-4010 Debrecen, PBox 105, Hungary}
}

\maketitle

\begin{abstract}
We present a systematic comparison of different approaches for the modeling of $t\bar{t}\gamma$ final states with leptonic decays at the LHC. On the one hand, we consider a complete calculation at NLO QCD accuracy which includes all resonant and non-resonant diagrams. On the other hand, we consider predictions in the narrow-width approximation with top quark decays modeled at various accuracies. In this way we quantify the impact of the off-shell effects in $t\bar{t}\gamma$ production. We also discuss the relative importance of double-, single- and non-resonant contributions in the complete calculation. Finally we investigate the fraction of isolated photons from decays of top quarks, which represent a background for measurements of anomalous $t\gamma$ couplings.
\end{abstract}


\section{Introduction}

Precision measurements of processes involving top quark pair production provide a unique opportunity for testing the Standard Model (SM) at the LHC. Using the full luminosity collected at Run II, associated production channels such as $t\bar{t}H$ or $t\bar{t}V \, (V=\gamma,Z,W^\pm)$ can be studied with some detail. Even though the latter have cross sections that are orders of magnitude smaller than the inclusive $t\bar{t}$ production, they add significantly to the study of top quark properties at the LHC. Particularly interesting is the analysis of $t\bar{t}\gamma$ production in connection with precision measurements of properties such as  top quark electric charge \cite{Baur:2001si} or charge asymmetries \cite{Aguilar-Saavedra:2014vta,Bergner:2018lgm}. Being a natural probe of the $t\gamma$ vertex, this process can contribute to shed light on possible effects of physics beyond the SM (see \textit{e.g.} \cite{Bylund:2016phk,Schulze:2016qas}). Clearly, precise SM predictions are a prerequisite to achieve all these goals. 

The present state-of-the-art description of $t\bar{t}\gamma$ is NLO. Both QCD and EW corrections have been calculated in the picture of stable top quarks \cite{PengFei:2009ph,PengFei:2011qg,Duan:2016qlc,Maltoni:2015ena}. Predictions based on the narrow-width approximation are available at NLO QCD accuracy, including radiative effects in top quark decays \cite{Melnikov:2011ta}. Results matched to parton showers are also available for on-shell top quarks \cite{Kardos:2014zba}. More recently, predictions for the dilepton channel based on a complete NLO QCD calculation have started to appear \cite{Bevilacqua:2018woc}. Different approaches of modeling $t\bar{t}\gamma$ events are currently being analysed for the measurement of inclusive and differential cross sections in the $e\mu$ channel at 13 TeV by the ATLAS collaboration \cite{ATLAS:2019gkg}.

The approach of a complete calculation provides the most realistic description for a wide range of observables and, without any doubt, should be used when possible. On the other hand, it is well known that such approach is often demanding computationally and results based on the narrow-width approximation are sufficiently accurate under certain conditions. The only way to critically assess the accuracy of approximate results is to perform systematic comparisons at the differential level. With this motivation at hand, we have performed a comparative study of various approaches for the modeling of $t\bar{t}\gamma$ from the viewpoint of a fixed-order calculation, focusing on the dilepton channel.  We report on the results of this work, as presented in \cite{Bevilacqua:2019quz}.

\section{Details of the calculation}

We study the process $pp \to e^+ \nu_e \mu^- \bar{\nu}_\mu b \bar{b} \, \gamma + X$ at NLO QCD accuracy, considering the LHC Run II energy of 13 TeV. The top quark mass is set to $m_t = 173.2$ GeV while all other fermions are treated as massless. We consider two different functional forms for the renormalization and factorization scales: $\mu_R = \mu_F = m_t/2$ and $\mu_R = \mu_F = H_T/4$, where
\begin{equation}
H_T = p_T(e^+) + p_T(\mu^-) + p_T(b_1) + p_T(b_2)  + p_T(\gamma) + p_T^{miss} \,,
\end{equation}
and $b_1,b_2$ denote $b$-jets. The first scale prescription is a common choice in various phenomenological studies while the second one, phase-space dependent, is our recommendation based on earlier studies \cite{Bevilacqua:2018woc}.  Scale uncertainties are estimated by varying the default values of the renormalization and factorization scales independently by a factor of 2 and taking the envelope of the resulting predictions. We consider the CT14 \cite{Dulat:2015mca}, MMHT14 \cite{Harland-Lang:2014zoa}, and NNPDF3.0 \cite{Ball:2014uwa} parton distribution functions (PDF) in accordance with the PDF4LHC recommendations for LHC Run II \cite{Butterworth:2015oua}. For further details of the computational setup we refer to our published work \cite{Bevilacqua:2019quz,Bevilacqua:2018woc}.

On the technical side, our results have been obtained with the help of the package \textsc{Helac-Nlo} \cite{Bevilacqua:2011xh}. Real-emission contributions are calculated with the Nagy-Soper scheme \cite{Bevilacqua:2013iha} and cross-checked with the Catani-Seymour scheme \cite{Catani:1996vz,Catani:2002hc}. In both cases we adopt a formulation valid for arbitrary helicity eigenstates of the external partons \cite{Czakon:2009ss}. Phase-space integration is performed with \textsc{Kaleu} \cite{vanHameren:2010gg}.
Our results are available in the form of events in Les Houches Event File format \cite{Alwall:2006yp} or ROOT Ntuples \cite{Antcheva:2009zz} that might be directly used for experimental studies. Each event is stored with additional matrix-element and PDF information which allows on-the-fly reweighting for different scales and PDFs \cite{Bern:2013zja}. A newly developed tool, \textsc{Heplot}, can be used to obtain predictions for arbitrary infrared-safe observables and kinematical cuts from the Ntuples, together with full theoretical uncertainties stemming from scale and PDF variations. 

As already mentioned, a comprehensive analysis of various approaches for the modeling of $t\bar{t}\gamma$ final states is the main focus of this work. We will compare results from a full calculation against those based on the narrow-width approximation for top quarks and $W$ bosons. For ease of notation, we will denote the two approaches "full off-shell" and "NWA". In the former, all resonant and non-resonant Feynman diagrams, interferences and finite-width effects at the perturbative order $\mathcal{O}(\alpha^5 \alpha_s^3)$ are taken into account. In the latter only double-resonant contributions are retained, where top quarks and $W$ bosons are produced on-shell and decayed with full spin correlations. 
In order to facilitate more systematic comparisons we have extended the \textsc{Helac-Nlo} framework with the capability to perform calculations in full NWA. We summarize in the next section the main features of our implementation in comparison with other standard approaches.

\section{Narrow Width Approximation in \textsc{HELAC-NLO}}

The NWA offers a conceptually easy and powerful framework for computing processes characterized by the production of unstable resonances when the width ($\Gamma$) of such particles is small compared to their mass ($m$). The factorization of the cross section into production times decay is driven by the limit
\begin{equation}
\label{Eq:NWA}
\frac{1}{(p^2 - m^2)^2 + m^2\Gamma^2} \stackrel{\Gamma/m \to 0}{\longrightarrow} \frac{\pi}{m\Gamma} \, \delta(p^2-m^2) + \mathcal{O}(\frac{\Gamma}{m}) \,.
\end{equation}
Non-resonant contributions are systematically removed from the computation of scattering amplitudes in NWA. Such contributions are suppressed by the ratio $\Gamma/m$ for sufficiently inclusive observables \cite{Fadin:1993kt}, although they are well known to have a more prominent role in certain regions of phase space (see \textit{e.g.} \cite{Kauer:2001sp,Denner:2012mx}). In standard implementations of NWA, the amplitudes of the various production and decay subprocesses are computed separately and combined later. In order to preserve spin correlations, bookkeeping of matrix elements for different polarizations of decaying particles is required. The combinatorial burden increases with the number of unstable particles and with the number of sequential decays. We will name this approach \textit{bottom-up} by virtue of its feature of combining simpler building blocks. 

In our work we adopt a \textit{top-bottom} approach. Instead of computing the various subprocesses separately, we take the viewpoint of the fully decayed final state. Amplitudes are calculated using  standard recursive algorithms, with simple modifications which restrict the computation to resonant contributions. These modifications shall be accompanied by some change in the propagators. For resonant fermionic propagators, according to formula (\ref{Eq:NWA}), the change reads\footnote{the Dirac delta appearing in Eq.(\ref{Eq:NWA}) is absorbed  in the phase space.}
\begin{equation}
\frac{\slashed{p}_f + m_f}{(p_f^2 - m_f^2) + i m_f \Gamma_f} \quad {\longrightarrow} \quad (\slashed{p}_f + m_f) \sqrt{\frac{\pi}{m_f \, \Gamma_f}}  \,,
\label{Eq:res_propag}
\end{equation}
while for non-resonant propagators we have
\begin{equation}
\frac{\slashed{p}_f + m_f}{(p_f^2 - m_f^2) + i m_f \Gamma_f} \quad {\longrightarrow} \quad  \frac{\slashed{p}_f + m_f}{(p_f^2 - m_f^2)} \,.
\label{Eq:nonres_propag}
\end{equation}
We note that the numerator in Eq.(\ref{Eq:res_propag}) can be left unchanged because $(\slashed{p}_f + m_f) = \sum_{s=\pm} u(p_f,s)\bar{u}(p_f,s)$ in the on-shell limit. The treatment of heavy-boson propagators is analogue. The top-down approach has the advantage of avoiding bookkeeping issues, which sounds appealing for processes featuring multiple sequential decays. Let us remark, at this point, that the two approaches are completely equivalent. We prefer the top-down approach because it minimizes structural changes in the framework of \textsc{Helac-Nlo} while being rather simple to implement.  We can use highly optimized algorithms, as developed for the computation of processes such as off-shell $t\bar{t}+X$ $(X=\gamma,Z,j)$ \cite{Bevilacqua:2018woc,Bevilacqua:2015qha,Bevilacqua:2016jfk,Bevilacqua:2019cvp,Bevilacqua:2018dny}, for the efficient selection of resonant contributions.

Further subtleties appear when dealing with NLO calculations. From the point of view of the virtual corrections, our approach does not set additional complications other than the efficient selection of loop topologies corresponding to factorizable corrections. From the point of view of the real corrections, the subtraction of infrared divergences requires some attention. Indeed the radiation of an unresolved gluon from resonant, on-shell propagators generates additional divergencies which are absent in the off-shell case. To treat such divergences, a few modifications are required in the organization of the subtraction. The gluon radiation mentioned above can be part either of the production process (when the gluon is radiated by a top quark that gets on-shell after radiation and decays) or of the decay process (as initial-state radiation). In the first case, the problem can be treated using standard Catani-Seymour dipoles. It is sufficient to include the resonant top quarks in the list of final-state emitters and compute the corresponding dipoles for the cases of final-state and initial-state spectators. The phase space mapping is applied to the momentum of the top quark, as reconstructed from its decay products. The mapping is  propagated afterwards to the daughter particles (see Ref.\cite{Bevilacqua:2019quz} for more details). Because the resonant propagator implicitly sums over the polarizations of the top quark, it is not possible to use the polarized formulae of Ref.\cite{Czakon:2009ss}. Crucially, the divergence is of pure soft nature and as such it is independent of the gluon polarization. Thus the required subtraction term can be simply set equal to the standard, non-polarized Catani-Seymour dipole with an additional factor 1/2 to avoid double counting in polarization sums. Let us now consider the second case, namely gluon radiation from top quark decays. In this case we use the prescription introduced in \cite{Campbell:2004ch} and generalized in \cite{Melnikov:2011ta} to the case of radiative top quark decays. The formula available in the literature refers to the unpolarized case. We devised a simple extension of it to the case of polarized partons which, for massless $b$-quarks, reads
\begin{equation}
  \label{Eq:NWA_finalinitial}
\begin{split}
 & D\left(\,(p_t+p_g)^2, (p_b+p_g)^2,
   m_t^2, M_W^2 \,\right)_{\lambda\lambda^\prime\lambda_b\lambda_g}
 =  \\[0.2cm]
 & g^2 \mu^{2\epsilon} C_F \left[
\frac{1}{p_b \cdot p_g} \left( \frac{z^2}{(1-z)} +
\delta_{\lambda_b\lambda_g}(1 + z) \right) -
\frac{1}{2}\frac{m_t^2}{\left(p_t \cdot p_g\right)^2} \right]
\delta_{\lambda\lambda_b}\delta_{\lambda\lambda^\prime} \,.
\end{split}
\end{equation}
Here $\lambda_b, \lambda_g$ are the helicity eigenstates of the external $b$-quark and gluon respectively and $\lambda,
\lambda^\prime$ are the helicity eigenstates that enter the Born matrix element. To conclude this part, we observe that the changes described above do not affect the analytical structure of the integrated dipoles. We used the formulae already available in the literature without need of any change.

\section{Numerical results}

In this section we present selected results from our study at the LHC with 13 TeV. Events with exactly two b-jets, two charged leptons, one hard photon and missing $p_T$ in the final state are selected. The photon is required to be isolated according to the prescription of Ref. \cite{Frixione:1998jh}. The following kinematical cuts are imposed:
\begin{equation}
\begin{array}{lclcl}
p_{T,\,\ell}>30 ~{\rm GeV}  &&  p_{T,\,b}>40  ~{\rm GeV}  &&  p^{miss}_{T} >20~{\rm GeV}  \\ [0.2cm]
\Delta R_{\ell b} > 0.4 && \Delta R_{bb}>0.4  &&  \Delta R_{\ell \ell} > 0.4 \\ [0.2cm]
|y_\gamma|<2.5 &&  |y_\ell|<2.5  &&  |y_b|<2.5 \,, 
\end{array}
\end{equation}
where $b$, $\ell$ denote respectively any $b$-jet and charged lepton. Furthermore, we require for the photon $p_{T,\gamma} > 25$ GeV, $\vert y_\gamma \vert < 2.5$ and $\Delta R_{\ell \gamma}>0.4$.
Jets are defined using the anti-$k_T$ clustering algorithm \cite{Cacciari:2008gp}, setting $R = 0.4$ as resolution parameter. No restriction is imposed on the extra jet other than the condition that it must be separated from the isolated photon. 

Let us begin the discussion with an analysis of the impact of different modeling approaches on the integrated cross sections. In Table \ref{Tab:integrated} we report our findings as obtained for the two scale prescriptions and using CT14 PDFs. We compare the full off-shell result against NWA with different levels of accuracy, namely: (i) decays at NLO and photon radiation in both production and decays (full NWA); (ii) decays at NLO and photon radiation in  production only (NWA${}_{\rm \gamma-prod}$); (iii) decays at NLO and photon radiation in decays only (NWA${}_{\rm \gamma-decay}$); (iv) decays at LO and photon radiation in production only (NWA${}_{\rm LOdecay}$). From Table \ref{Tab:integrated} we evince that, at NLO,  contributions of photon radiation from production and from decays are quite balanced as they amount respectively to 57\% and 43\%, independently on the scale choice. An important fraction of isolated photons is thus radiated off the top quark decay products. This finding is consistent with earlier published studies  based on full NWA \cite{Melnikov:2011ta}, which focused on different decay channels and collider energies. Thus, including radiative effects in the modeling of top quark decays is mandatory for reliable estimates of fiducial cross sections. Even more so, we observe that the NWA${}_{\rm LOdecay}$ prediction badly underestimates the full NLO QCD result. From Table \ref{Tab:integrated} one can also see that the off-shell effects change the NLO cross section by less than 3\% independently of the scale choice. This is consistent with expectations driven by the ratio $\Gamma_t/m_t \approx 0.8 \%$ and confirms once more that the full NWA does an excellent job for sufficiently inclusive observables.

\begin{figure}[t!hb]
\centerline{
\includegraphics[width=0.5\textwidth]{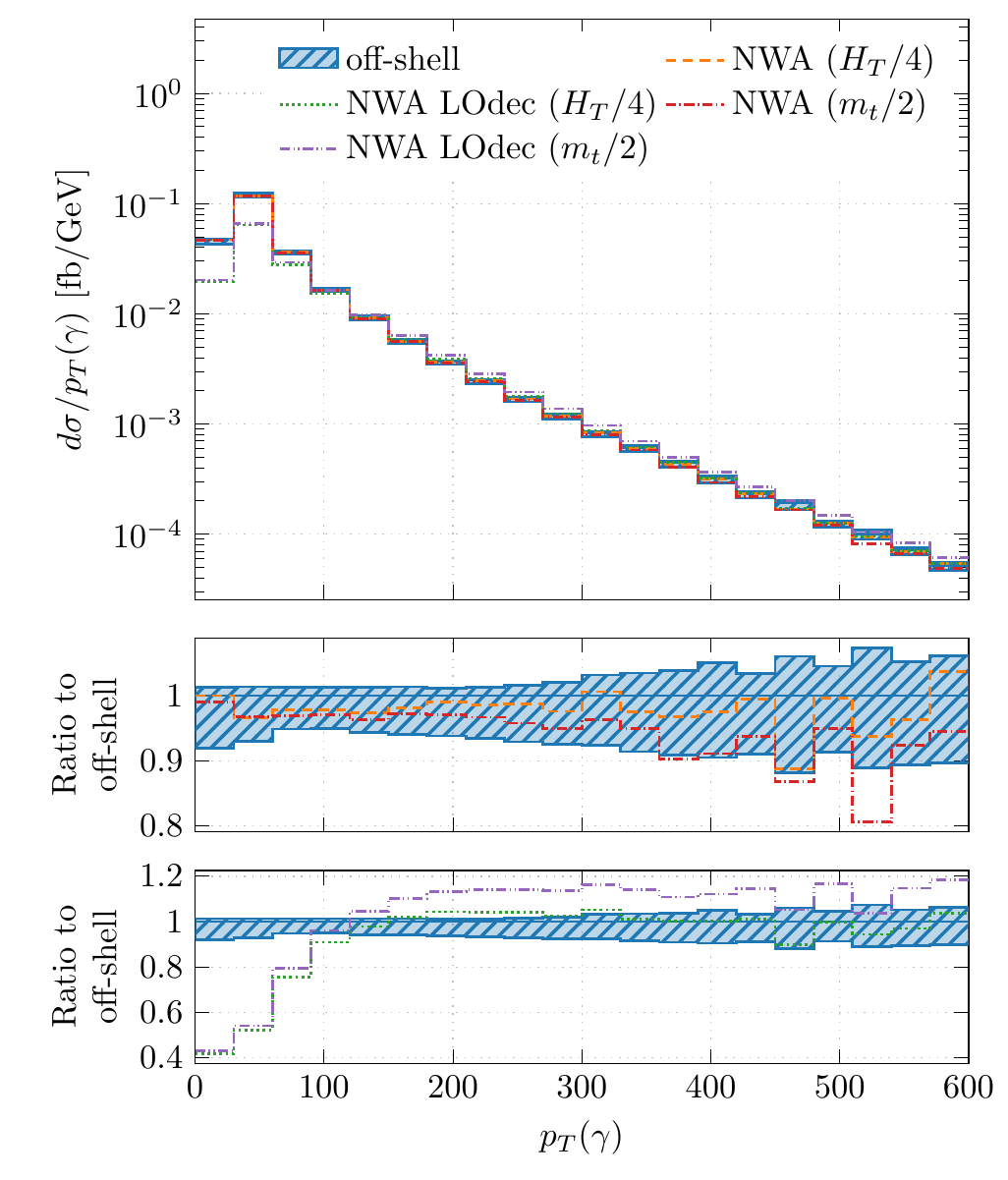}
\includegraphics[width=0.5\textwidth]{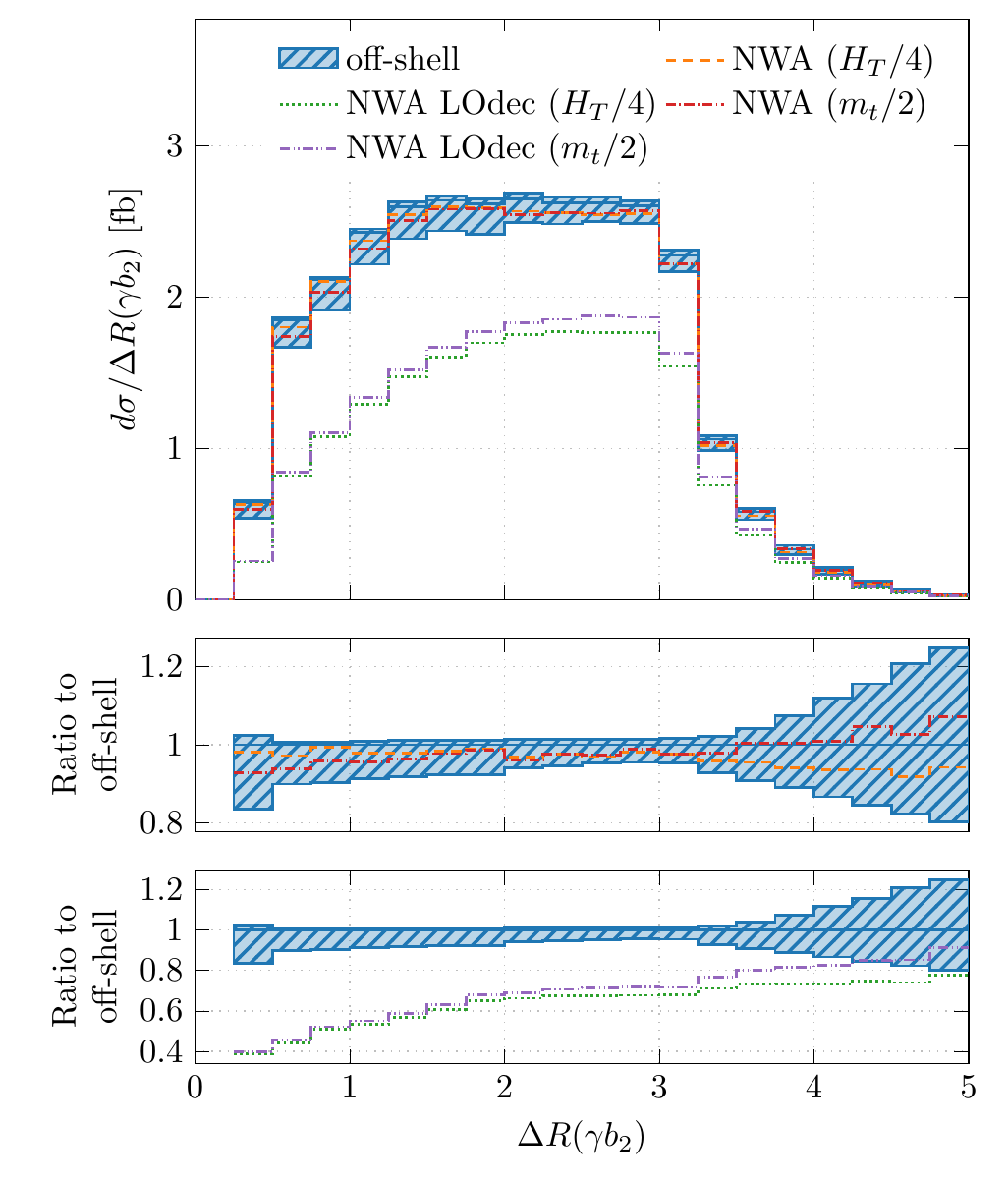}
}
\medskip
\centerline{
\includegraphics[width=0.5\textwidth]{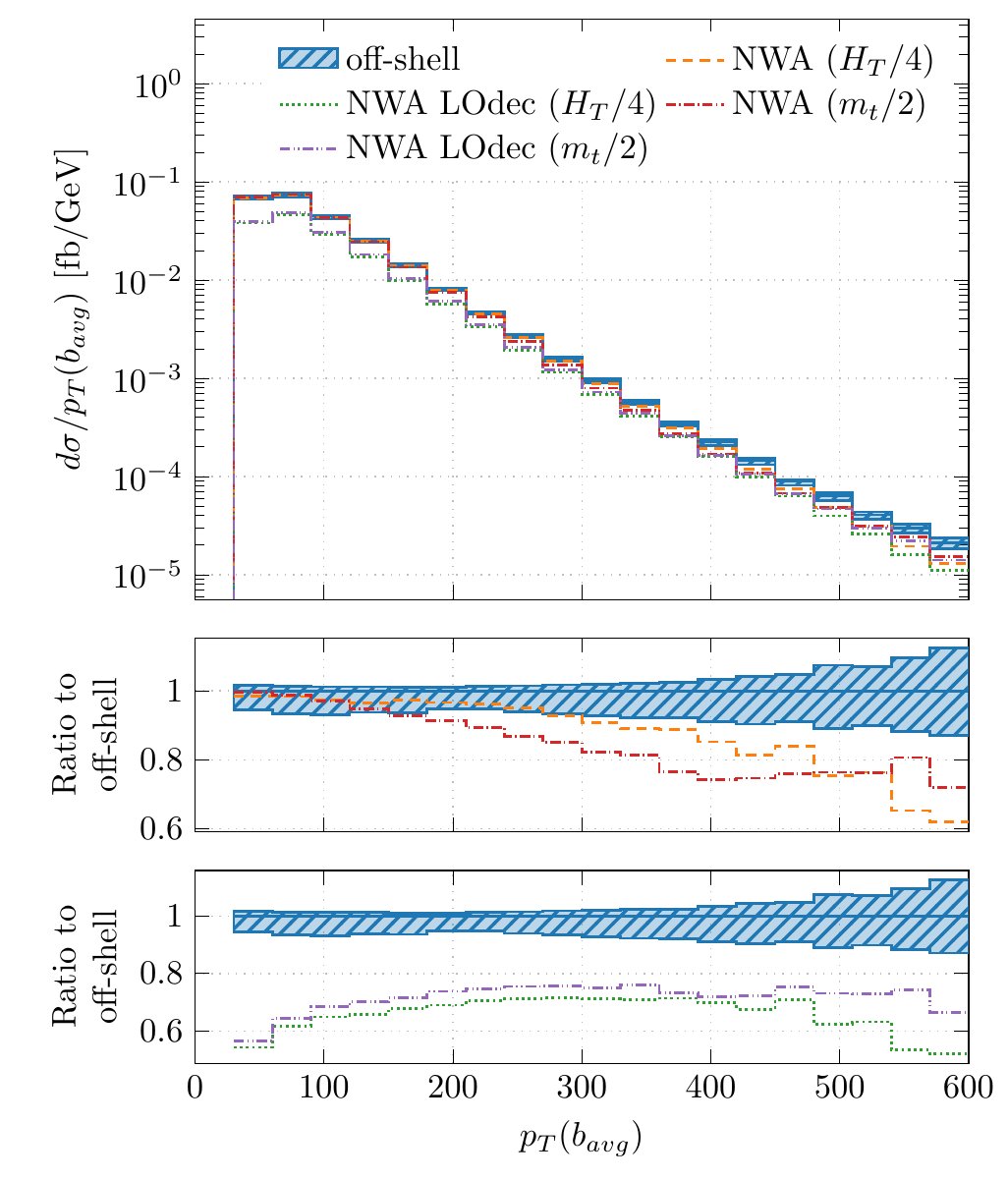}
\includegraphics[width=0.5\textwidth]{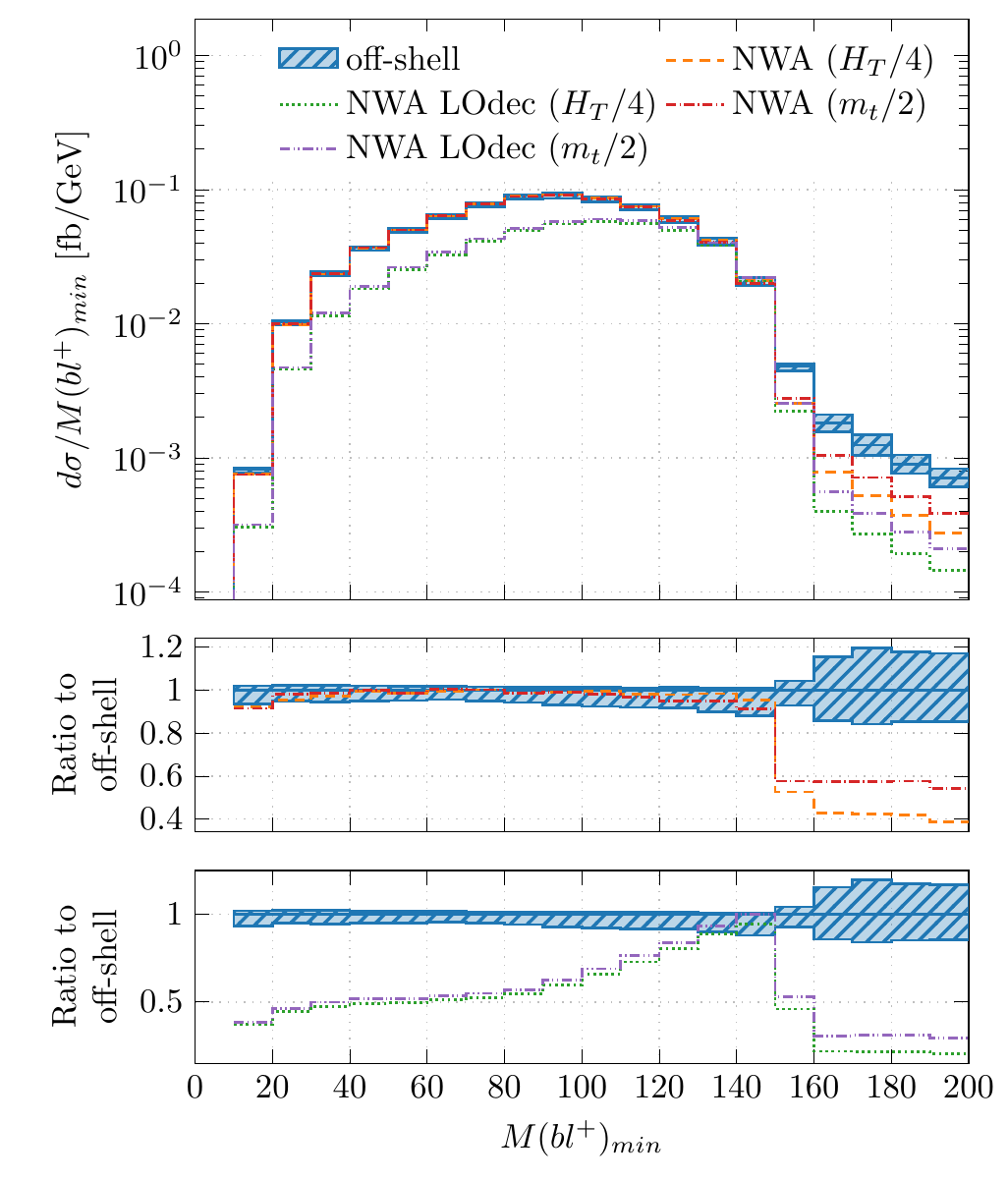}
}
\caption{ \label{Fig:compare_offshell} Differential cross sections for $pp\to e^+\nu_e \mu^- \bar{\nu}_\mu b \bar{b} \gamma +X$ as a function of $p_T(\gamma)$, $\Delta R(\gamma b_2)$, $p_T(b_{avg})$ and $M(b l^+)_{min}$ (defined in the text). \textit{Upper panels}: absolute NLO QCD predictions. \textit{Middle panels}: ratio between full NWA and off-shell. \textit{Lower panels}: ratio between NWA${}_{\rm LOdecay}$ and off-shell. The off-shell prediction is based on the scale choice $mu_R = mu_F = HT/4$. All results are based on CT14 PDFs. The uncertainty bands refer to the off-shell calculation with default scale $H_T/4$.}
\end{figure}

Let us now take a more exclusive point of view and check some differential cross sections. Figure \ref{Fig:compare_offshell} shows four distributions of phenomenological interest: the transverse momentum of the photon ($p_T(\gamma)$), the $\Delta R$ separation between the photon and the softest $b$-jet ($\Delta R(\gamma b_2)$), the average $p_T$ of the $b$-jets  ($p_T(b_{avg})$) and finally the minimum invariant mass between the positively charged lepton and the $b$-jets ($M(b l^+)_{min}$). The first two observables are well known for being sensitive to physics beyond the SM, while the latter has been widely investigated for top quark mass measurements in the $t\bar{t}(j)$ channel. The plots show results for the off-shell, full NWA and NWA${}_{\rm LOdecay}$ cases. The uncertainty band refers to the most accurate prediction, \textit{i.e.} the off-shell calculation. The accuracy of NWA is questionable where the NWA curves do not fit well within the uncertainty bands. As shown in Figure \ref{Fig:compare_offshell}, different observables have different behaviors: for $p_T(\gamma)$ and $\Delta R(\gamma b_2)$ the full NWA approach is accurate in the whole observed range, on the other hand for $p_T(b_{avg})$ and $M(b l^+)_{min}$ there are visible discrepancies in tails. We note again that predictions based on NWA${}_{\rm LOdecay}$ do not adequately describe the process.

\begin{figure}[h!tb]
\centerline{
\includegraphics[width=0.5\textwidth]{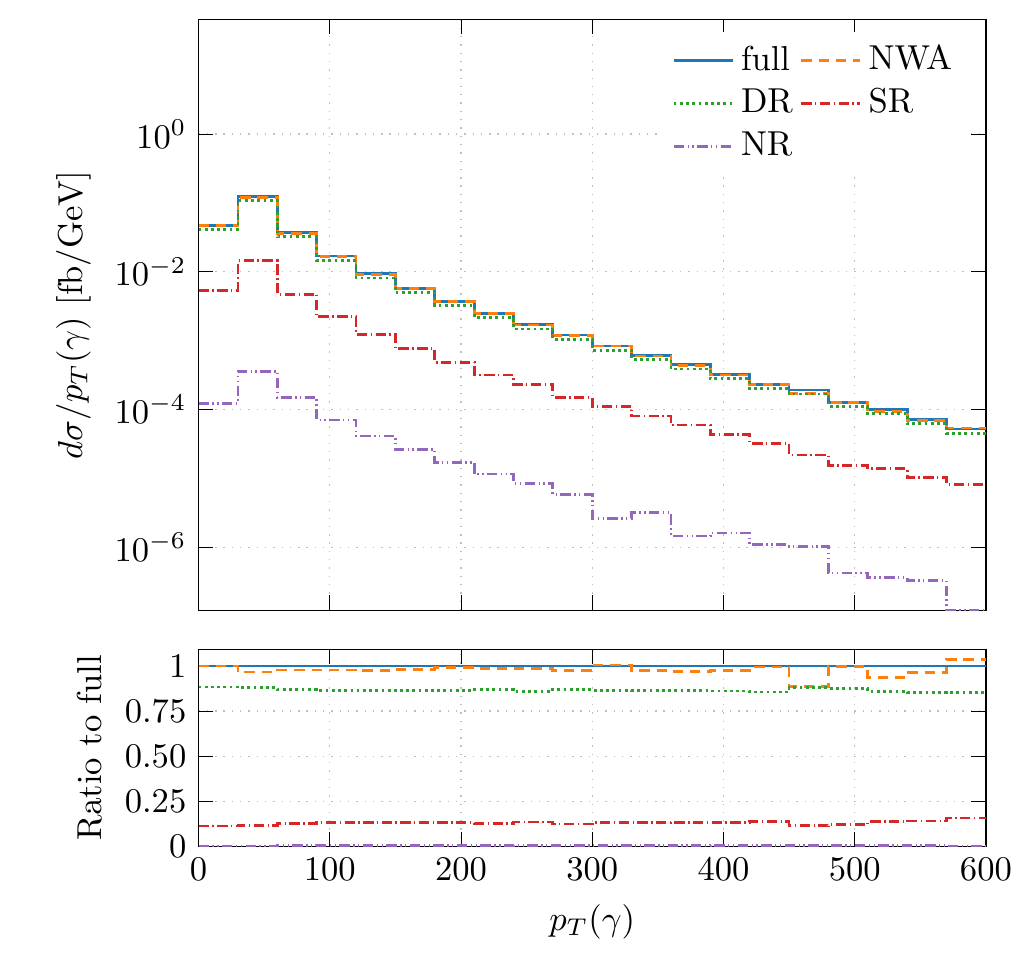}
\includegraphics[width=0.5\textwidth]{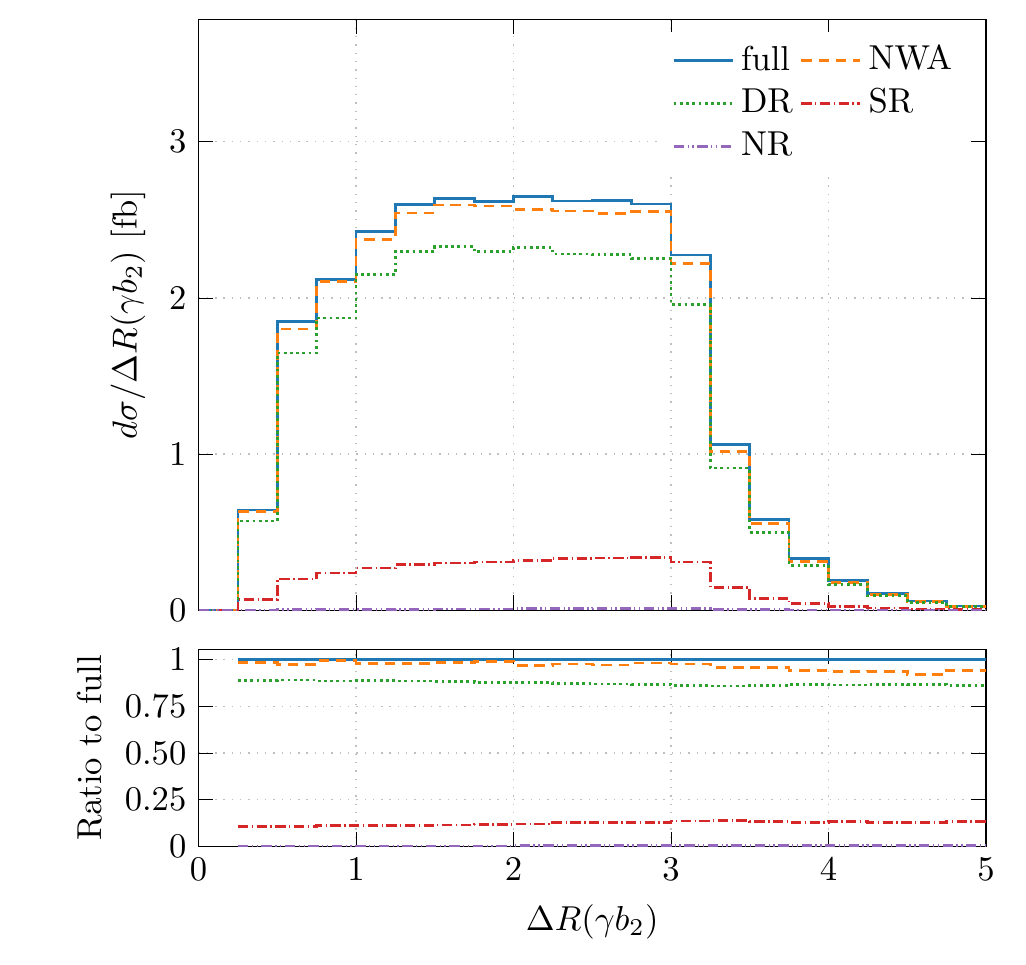}
}
\centerline{
\includegraphics[width=0.5\textwidth]{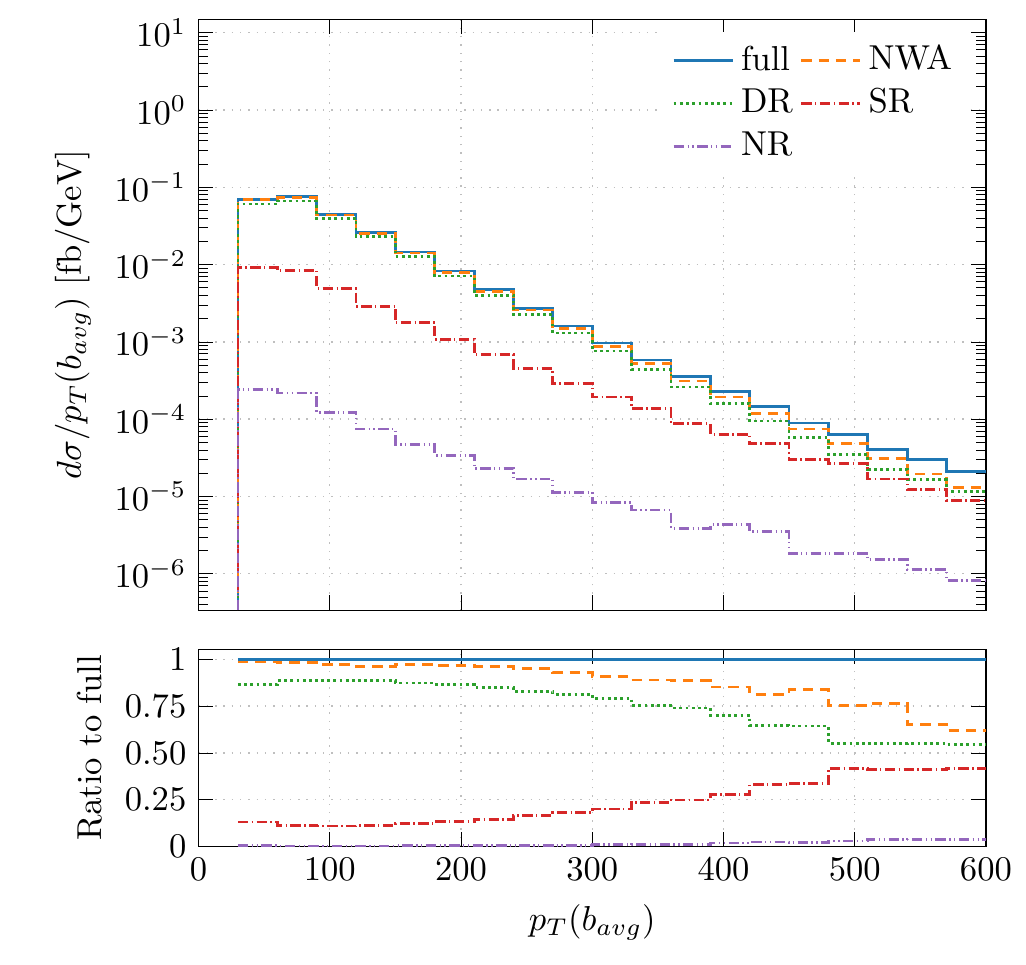}
\includegraphics[width=0.5\textwidth]{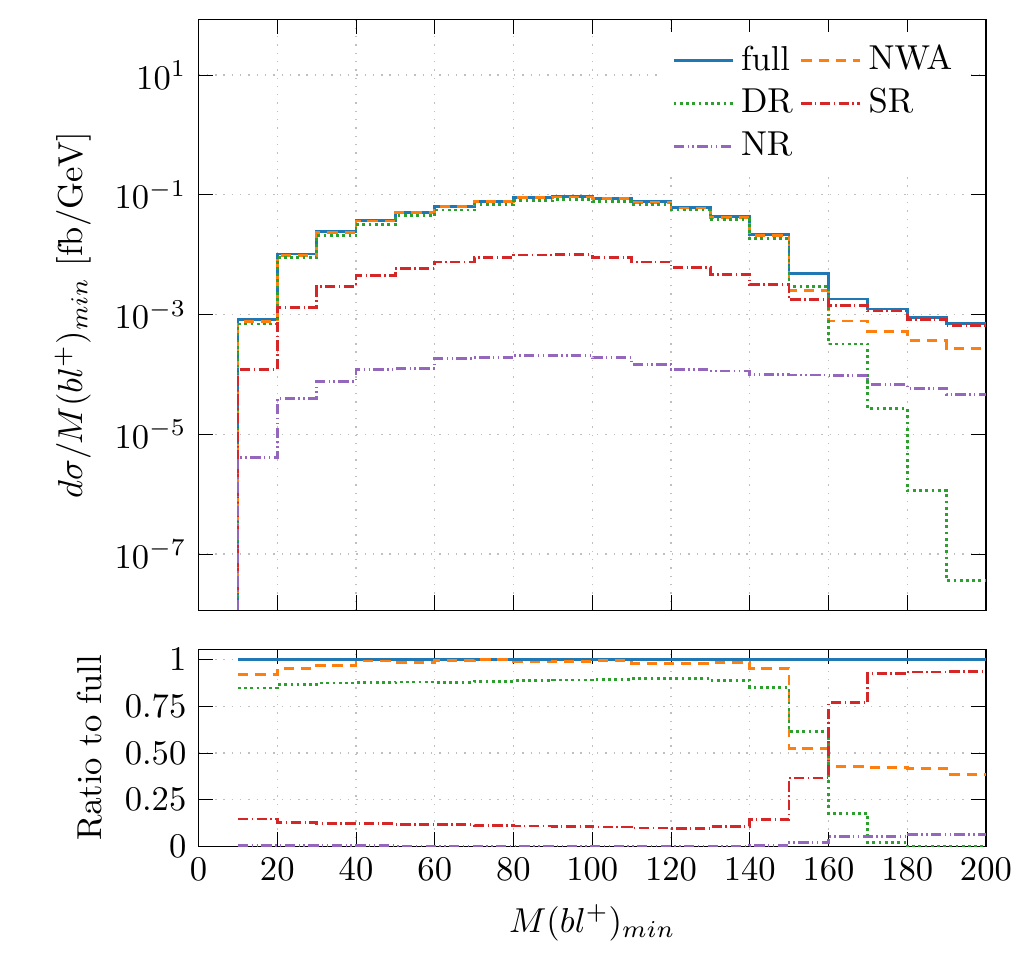}
}
\caption{ \label{Fig:compare_DR_SR_NR} Differential cross sections for $pp\to e^+\nu_e \mu^- \bar{\nu}_\mu b \bar{b} \gamma +X$ as a function of $p_T(\gamma)$, $\Delta R(\gamma b_2)$, $p_T(b_{avg})$ and $M(b l^+)_{min}$ (defined in the text). \textit{Upper panels}: absolute NLO QCD predictions for full off-shell as well as for double-, single- and non-resonant contributions (respectively DR, SR, NR). \textit{Lower panels}: ratio between DR/SR/NR and off-shell predictions. Results are based on the scale choice $mu_R = mu_F = HT/4$ and on CT14 PDFs.}
\end{figure}

To understand better why some observables are more sensitive to off-shell effects than others, it is helpful to investigate the relative importance of double-, single- and non-resonant contributions (denoted DR, SR and NR for brevity) in the full calculation. These are extracted with a selection procedure over the fiducial phase space which generalizes the method introduced in Ref. \cite{Kauer:2001sp}. The procedure can be sketched as follows: for any event, (i) we identify the most likely set of daughter particles from top quark decays and reconstruct $t$ and $\bar{t}$ invariant masses, then (ii) we check how much the reconstructed invariant masses differ  from the nominal mass, $m_t$. If the difference lies within a predefined window, the $t(\bar{t})$ quark is considered resonant, otherwise it is tagged as non-resonant.  Further details on this procedure can be found in Ref. \cite{Bevilacqua:2019quz}. It is clear that the partition into DR, SR and NR contributions is somewhat arbitrary in that it depends on the size of the window (we set it to be $15\,\Gamma_t \approx 20$ GeV). Yet, it is helpful to get an idea of the relative importance of the various contributions in different phase space regions. Our findings are reported in Figure \ref{Fig:compare_DR_SR_NR}, where we consider the same observables of Figure \ref{Fig:compare_offshell}. 
For the observables which proved to be less sensitive to the off-shell effects, DR contributions are nearly constant and dominant everywhere. In the other cases, we observe a correspondence between enhanced sensitivity to off-shell effects and increasing importance of SR contributions. NR contributions are extremely small in size everywhere.

\begin{figure}[h!tb]
\centerline{
\includegraphics[width=0.5\textwidth]{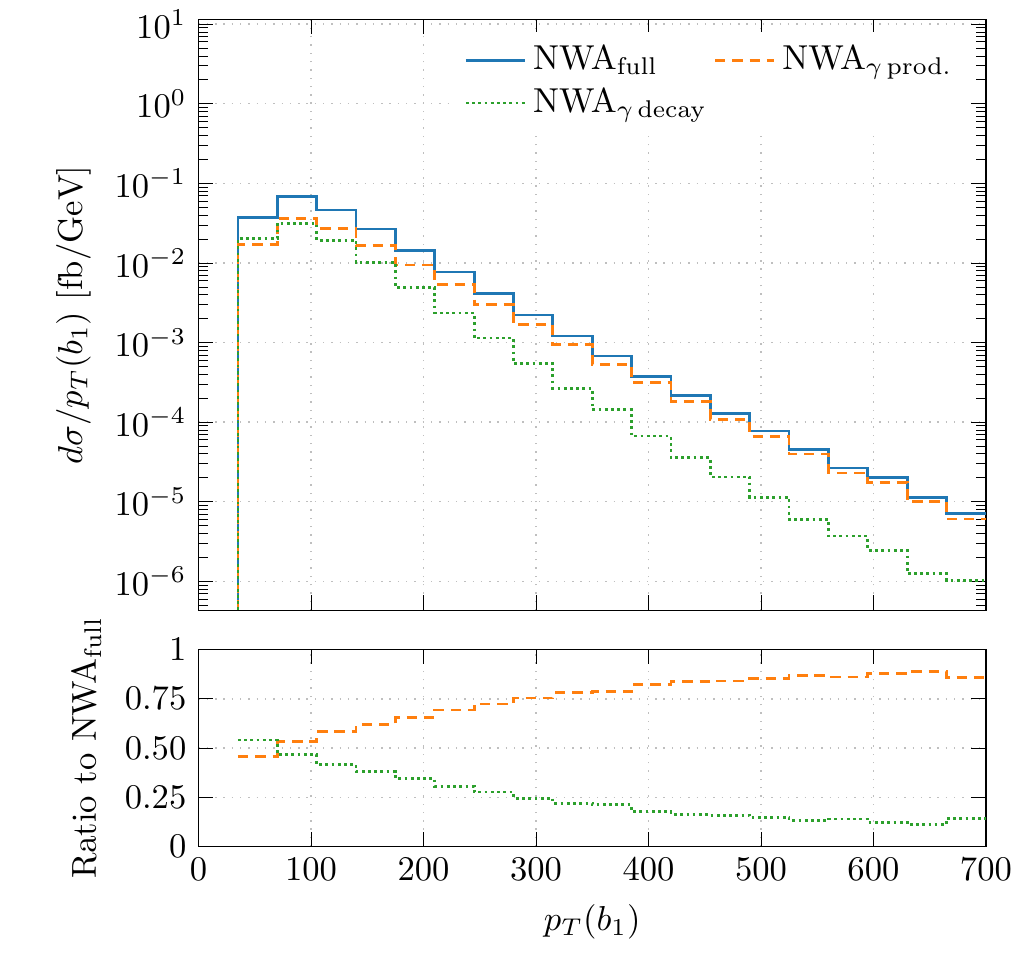}
\includegraphics[width=0.5\textwidth]{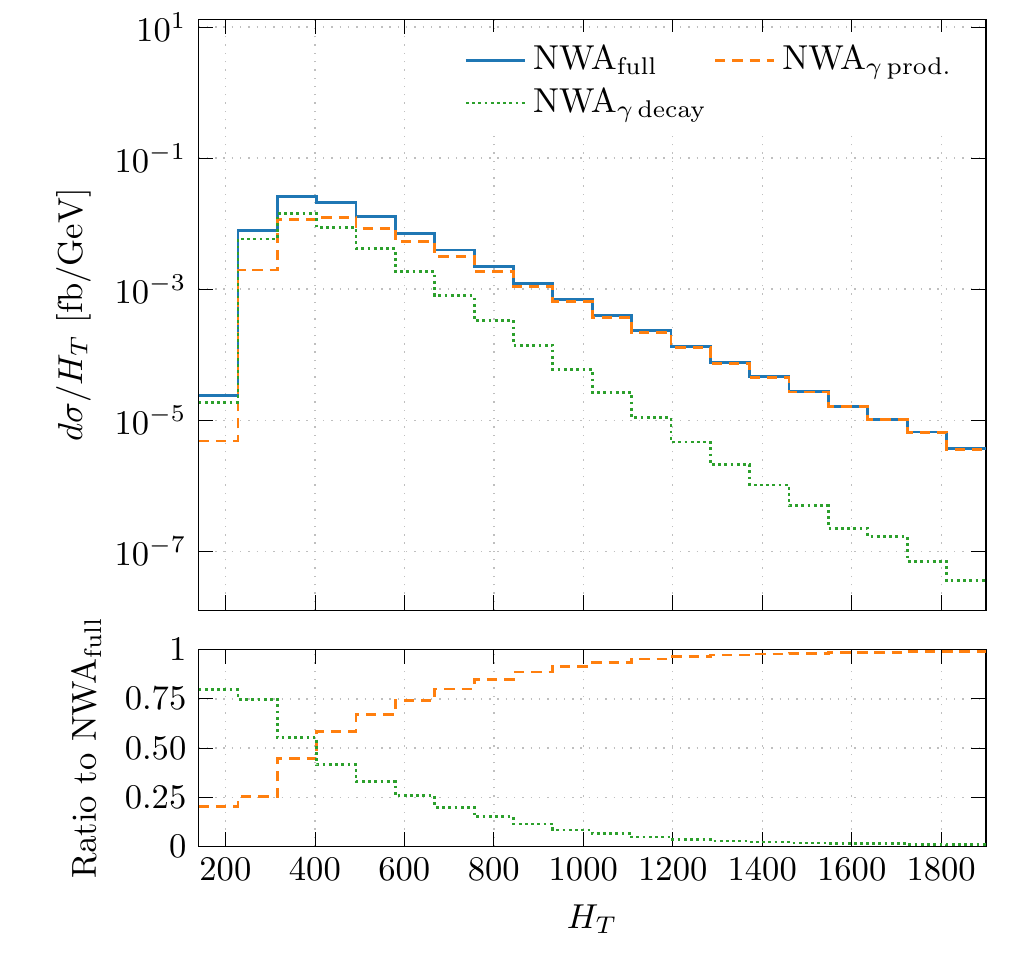}
}
\medskip
\centerline{
\includegraphics[width=0.5\textwidth]{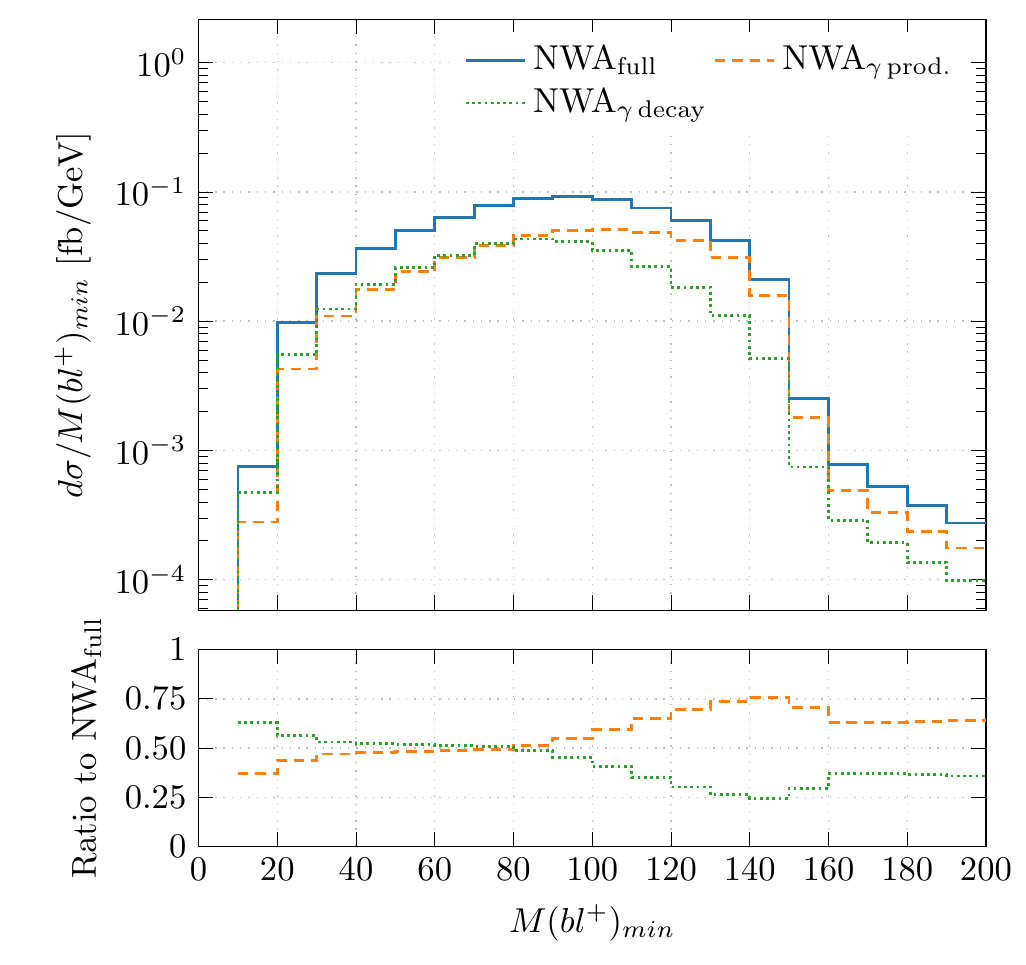}
\includegraphics[width=0.5\textwidth]{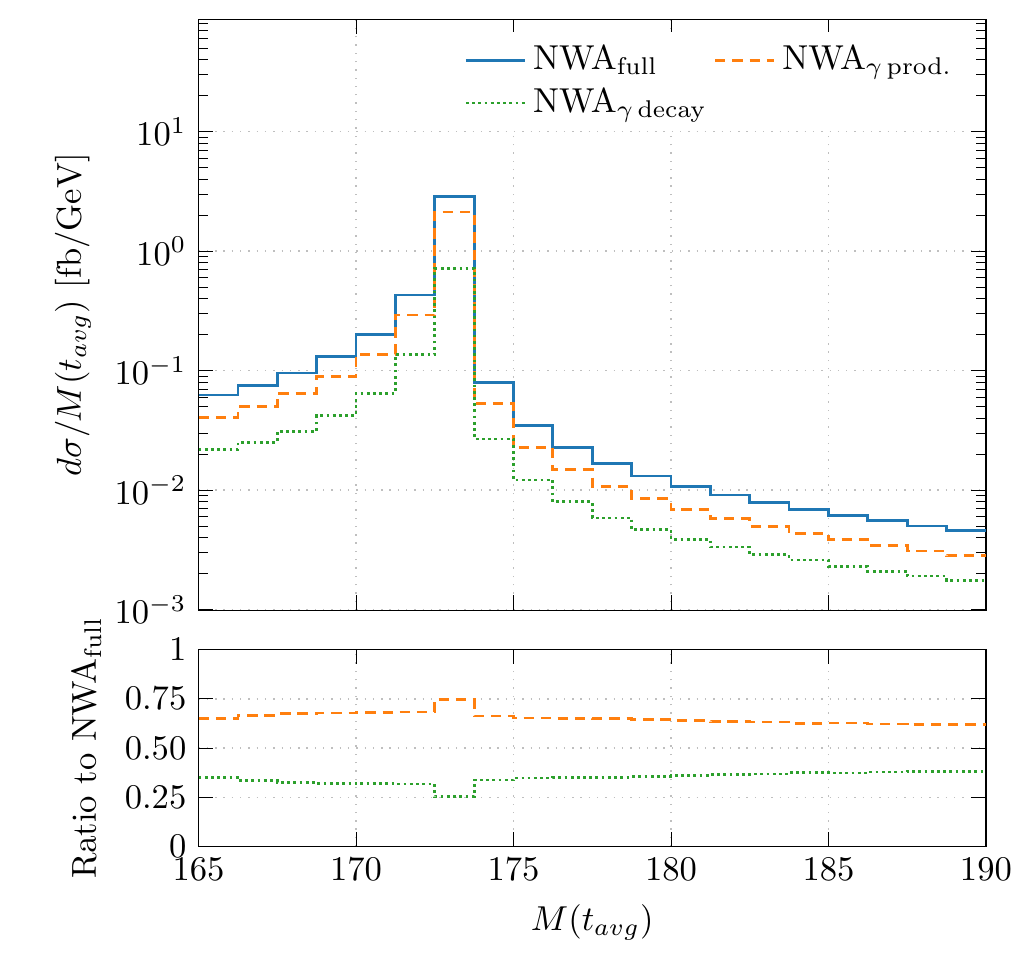}
}
\caption{ \label{Fig:GammaContrib} Differential cross sections for $pp\to e^+\nu_e \mu^- \bar{\nu}_\mu b \bar{b} \gamma +X$ as a function of $p_T(b_1)$, $H_T$, $M(b l^+)_{min}$ and $M(t_{avg})$ (defined in the text). \textit{Upper panels}: absolute NLO QCD predictions for full NWA as well as for NWA with photon radiation from production and from decays (respectively NWA${}_{\rm \gamma\,prod}$ and NWA${}_{\rm \gamma\,decay}$). \textit{Lower panels}: ratio between NWA${}_{\rm \gamma\,prod/decay}$ and full NWA predictions. Results are based on the scale choice $mu_R = mu_F = HT/4$ and on CT14 PDFs. }
\end{figure}

We conclude the discussion with a differential analysis of the fraction of events where the photon is radiated either in production or in  decays. The distinction between these contributions is well defined in NWA, where there is no cross talk between production and decay subprocesses. In the picture of a full calculation, non-factorizable and interference effects make such a net distinction impossible. 
One can suppress either contribution by use of suitable kinematical cuts, the design of which gets important feedback from NWA \cite{Melnikov:2011ta}.
Figure \ref{Fig:GammaContrib} shows four distributions, namely the $p_T$ of the hardest $b$-jet ($p_T(b_1)$), the average invariant mass of the reconstructed top quarks ($M(t_{avg})$) as well as the already introduced $M(b l^+)_{min}$ and $H_T$. They represent interesting cases of dimensionful observables with variable behaviours. In the case of $M(t_{avg})$, the relative contributions of NWA${}_{\rm \gamma\,prod}$ and NWA${}_{\rm \gamma\,decay}$  are rather constant and the first one dominates over the whole range, while for $M(b l^+)_{min}$ the two contributions have comparable size in some range. Finally, $p_T(b_1)$ and $H_T$ exhibit more distinct regions of influence, with photons from production (decay) dominating the hard (soft) part of the spectrum. These findings can be used  to further develop selection criteria to reduce the contribution of hard photons from top quark decays, which constitute a background for measurements of anomalous couplings in the $t\gamma$ vertex.

\begin{table}[h!]
  \begin{center}
\begin{tabular}{lcc}
  \hline \hline
  &&\\
  \textsc{Modeling Approach} & $\sigma^{\rm LO}$ [{\rm fb}]
                              & $\sigma^{\rm NLO}$ [{\rm  fb}]
  \\[0.2cm]
  \hline \hline
  &&\\
 full off-shell $(\mu_0=m_t/2)$ & ${8.28}^{+2.92\, (35\%)}_{-2.01\, (24\%)}$
 & ${7.44}^{+0.07\,(1\%)}_{-1.04\, (14\%)}$  \\[0.2cm]
 full off-shell $(\mu_0=H_T/4)$ & ${7.32}^{+2.45\, (33\%)}_{-1.71\, (23\%)}$
 & ${7.50}^{+0.11\,(1\%)}_{-0.45\, (6\%)}$  \\[0.2cm]\hline\hline
 &&\\ 
        NWA  $(\mu_0=m_t/2)$ &
                               ${8.08}^{+2.84\,(35\%)}_{-1.96\,(24\%)}$
                              & ${7.28}_{-0.03\,(0.4\%)}^{-0.99\,(13\%)}$   \\[0.2cm]
        NWA  $(\mu_0=H_T/4)$ &
                               ${7.18}^{+2.39\,(33\%)}_{-1.68\,(23\%)}$
                              & ${7.33}_{-0.24\,(3.3\%)}^{-0.43\,(5.9\%)}$
  \\[0.2cm]
   \hline \hline
  &&\\
  NWA${}_{\gamma-{\rm prod}}$ $(\mu_0=m_t/2)$
  & ${4.52}^{+1.63\,(36\%)}_{-1.11\,(24\%)}$  &
                                                ${4.13}_{-0.05\,(1.2\%)}^{-0.53\,(13\%)}$
  \\[0.2cm]
  NWA${}_{\gamma-{\rm prod}}$ $(\mu_0=H_T/4)$
  & ${3.85}^{+1.29\,(33\%)}_{-0.90\,(23\%)}$  &
    ${4.15}_{-0.21\,(5.1\%)}^{-0.12\,(2.3\%)}$   \\[0.2cm]
   \hline \hline
  &&\\
  NWA${}_{\gamma-{\rm decay}}$ $(\mu_0=m_t/2)$
  & ${3.56}^{+1.20\,(34\%)}_{-0.85\,(24\%)}$  &
                                                ${3.15}_{+0.03\,(0.9\%)}^{-0.46\,(15\%)}$
  \\[0.2cm]
  NWA${}_{\gamma-{\rm decay}}$ $(\mu_0=H_T/4)$
  & ${3.33}^{+1.10\,(33\%)}_{-0.77\,(23\%)}$  &
                                                ${3.18}^{-0.31\,(9.7\%)}_{-0.03\,(0.9\%)}$
  \\[0.2cm]
   \hline \hline
  &&\\
  NWA${}_{\rm LOdecay}$ $(\mu_0=m_t/2)$  &
                             & ${4.85}^{+0.26\,(5.4\%)}_{-0.48\,(9.9\%)}$   \\[0.2cm]
        NWA${}_{\rm LOdecay}$ $(\mu_0=H_T/4)$  &
                                                 & ${4.63}^{+0.44\,(9.5\%)}_{-0.52\,(11\%)}$\\[0.2cm]
 \hline     \hline                                             
\end{tabular}
\end{center}
\caption{\label{Tab:integrated}\it Integrated cross sections for $pp\to e^+\nu_e \mu^- \bar{\nu}_\mu b \bar{b} \gamma +X$ at  $\sqrt{s}=13$ TeV. Results of the off-shell calculation as well as of various approaches for the modelling of top quark decays in NWA are shown. The reported errors refer to uncertainties stemming from scale variation. All results refer to the CT14 PDF set.}
\label{tab:moddeling}
\end{table}

\section{Conclusions}

We have presented a comparative study of various approaches of modeling $t\bar{t}\gamma$ production in the dilepton channel at the LHC. Comparing the fully realistic description as given by a complete calculation with the one provided by the NWA, we have quantified for the first time at NLO QCD the size of the off-shell effects.
We discussed examples of differential cross sections that are relatively insensitive to off-shell effects and can be safely described by use of full NWA. At the same time we presented cases where the latter effects are visibly enhanced and a complete calculation should be used instead. Furthermore, we have shown that without including radiative effects in top quark decays (QCD corrections and photon radiation), the NWA does not adequately describe the process at hand.
On the technical side, all the results presented in this study have been computed with \textsc{Helac-Nlo}. The code has been extended to perform automated calculations in full NWA. Such automation will open the road to achieve predictions in full NWA for even more complex processes, such as $t\bar{t}b\bar{b}$, $t\bar{t}jj$ and $t\bar{t}t\bar{t}$ production.

\section*{Acknowledgements}
The research of G.~B. was supported by grant K 125105 of the National Research, Development and Innovation Office in Hungary.


\begin{thebibliography}{99}

\bibitem{Baur:2001si}
  U.~Baur, M.~Buice and L.~H.~Orr,
  \href{https://journals.aps.org/prd/abstract/10.1103/PhysRevD.64.094019}{Phys.\ Rev.\ D {\bf 64} (2001) 094019}    

\bibitem{Aguilar-Saavedra:2014vta}
  J.~A.~Aguilar-Saavedra, E.~Alvarez, A.~Juste and F.~Rubbo,
  \href{https://link.springer.com/article/10.1007/JHEP04(2014)188}{JHEP {\bf 1404} (2014) 188}    

\bibitem{Bergner:2018lgm}
  J.~Bergner and M.~Schulze,
  \href{https://dx.doi.org/10.1140/epjc/s10052-019-6707-6}{Eur.\ Phys.\ J.\ C {\bf 79} (2019) no.3,  189}     

\bibitem{Bylund:2016phk}
  O.~Bessidskaia Bylund, F.~Maltoni, I.~Tsinikos, E.~Vryonidou and C.~Zhang,
  \href{https://doi.org/10.1007/JHEP05(2016)052}{JHEP {\bf 1605} (2016) 052}     
  
\bibitem{Schulze:2016qas}
  M.~Schulze and Y.~Soreq,
  \href{https://link.springer.com/article/10.1140\%2Fepjc\%2Fs10052-016-4263-x}{Eur.\ Phys.\ J.\ C {\bf 76} (2016) no.8,  466}      

\bibitem{PengFei:2009ph}
  P.~F.~Duan, W.~G.~Ma, R.~Y.~Zhang, L.~Han, L.~Guo and S.~M.~Wang,
  \href{https://doi.org/10.1103/PhysRevD.80.014022}{Phys.\ Rev.\ D {\bf 80} (2009) 014022}     

\bibitem{PengFei:2011qg}
  P.~F.~Duan, R.~Y.~Zhang, W.~G.~Ma, L.~Han, L.~Guo and S.~M.~Wang,
  \href{https://dx.doi.org/10.1088/0256-307X/28/11/111401}{Chin.\ Phys.\ Lett.\  {\bf 28} (2011) 111401}    

\bibitem{Duan:2016qlc}
  P.~F.~Duan, Y.~Zhang, Y.~Wang, M.~Song and G.~Li,
  \href{https://www.sciencedirect.com/science/article/pii/S0370269317300035?via\%3Dihub}{Phys.\ Lett.\ B {\bf 766} (2017) 102}     

\bibitem{Maltoni:2015ena}
  F.~Maltoni, D.~Pagani and I.~Tsinikos,
  \href{https://link.springer.com/article/10.1007\%2FJHEP02\%282016\%29113}{JHEP {\bf 1602} (2016) 113}      

 \bibitem{Melnikov:2011ta}
  K.~Melnikov, M.~Schulze and A.~Scharf,
  \href{https://journals.aps.org/prd/abstract/10.1103/PhysRevD.83.074013}{Phys.\ Rev.\ D {\bf 83} (2011) 074013}     
    
  \bibitem{Kardos:2014zba}
  A.~Kardos and Z.~Trocsanyi,
  \href{https://link.springer.com/article/10.1007\%2FJHEP05\%282015\%29090}{JHEP {\bf 1505} (2015) 090}   

\bibitem{Bevilacqua:2018woc}
  G.~Bevilacqua, H.~B.~Hartanto, M.~Kraus, T.~Weber and M.~Worek,
  \href{https://link.springer.com/article/10.1007\%2FJHEP10\%282018\%29158}{JHEP {\bf 1810} (2018) 158}    

\bibitem{ATLAS:2019gkg}
   ATLAS Collaboration,
  \href{https://cds.cern.ch/record/2690350}{ATLAS-CONF-2019-042}  (2019).

\bibitem{Bevilacqua:2019quz}
  G.~Bevilacqua, H.~B.~Hartanto, M.~Kraus, T.~Weber and M.~Worek,
   \href{https://link.springer.com/article/10.1007\%2FJHEP03\%282020\%29154}{JHEP {\bf 03} (2020) 154}     

\bibitem{Dulat:2015mca}
  S.~Dulat {\it et al.},
  \href{https://journals.aps.org/prd/abstract/10.1103/PhysRevD.93.033006}{Phys.\ Rev.\ D {\bf 93} (2016) no.3,  033006}   

\bibitem{Harland-Lang:2014zoa}
  L.~A.~Harland-Lang \textit{et al.},
  \href{https://dx.doi.org/10.1140/epjc/s10052-015-3397-6}{Eur.\ Phys.\ J.\ C {\bf 75} (2015) no.5,  204}     

\bibitem{Ball:2014uwa}
  R.~D.~Ball {\it et al.} [NNPDF Collaboration],
  \href{https://link.springer.com/article/10.1007\%2FJHEP04\%282015\%29040}{JHEP {\bf 1504} (2015) 040}     

\bibitem{Butterworth:2015oua}
  J.~Butterworth {\it et al.},
  \href{https://dx.doi.org/10.1088/0954-3899/43/2/023001}{J.\ Phys.\ G {\bf 43} (2016) 023001}     
  
  \bibitem{Bevilacqua:2011xh}
  G.~Bevilacqua, M.~Czakon, M.~V.~Garzelli, A.~van Hameren, A.~Kardos, C.~G.~Papadopoulos, R.~Pittau and M.~Worek,
  \href{https://www.sciencedirect.com/science/article/pii/S0010465512003761?via\%3Dihub}{Comput.\ Phys.\ Commun.\  {\bf 184} (2013) 986}     
  
 \bibitem{Bevilacqua:2013iha}
  G.~Bevilacqua, M.~Czakon, M.~Kubocz and M.~Worek,
  \href{https://link.springer.com/article/10.1007\%2FJHEP10\%282013\%29204}{JHEP {\bf 1310} (2013) 204}     
  
\bibitem{Catani:1996vz}
  S.~Catani and M.~H.~Seymour,
  \href{https://www.sciencedirect.com/science/article/pii/S0550321396005895?via\%3Dihub}{Nucl.\ Phys.\ B {\bf 485} (1997) 291}
  Erratum: [Nucl.\ Phys.\ B {\bf 510} (1998) 503]    
  
\bibitem{Catani:2002hc}
 S.~Catani, S.~Dittmaier, M.~H.~Seymour and Z.~Trocsanyi,
 \href{https://www.sciencedirect.com/science/article/pii/S0550321302000986?via\%3Dihub}{Nucl.\ Phys.\ B {\bf 627} (2002) 189}    

\bibitem{Czakon:2009ss}
  M.~Czakon, C.~G.~Papadopoulos and M.~Worek,
  \href{https://dx.doi.org/10.1088/1126-6708/2009/08/085}{JHEP {\bf 0908} (2009) 085}    

\bibitem{vanHameren:2010gg}
  A.~van Hameren,
  \href{https://arxiv.org/abs/1003.4953}{arXiv:1003.4953 [hep-ph]}.

\bibitem{Alwall:2006yp}
  J.~Alwall {\it et al.},
  \href{https://www.sciencedirect.com/science/article/pii/S0010465506004164?via\%3Dihub}{Comput.\ Phys.\ Commun.\  {\bf 176} (2007) 300}    

\bibitem{Antcheva:2009zz}
  I.~Antcheva {\it et al.},
  \href{https://www.sciencedirect.com/science/article/pii/S0010465509002550?via\%3Dihub}{Comput.\ Phys.\ Commun.\  {\bf 180} (2009) 2499}    
  
\bibitem{Bern:2013zja}
  Z.~Bern \textit{et al.},
  \href{https://www.sciencedirect.com/science/article/pii/S0010465514000241?via\%3Dihub}{Comput.\ Phys.\ Commun.\  {\bf 185} (2014) 1443}     
  
\bibitem{Frixione:1998jh}
  S.~Frixione,
  \href{https://www.sciencedirect.com/science/article/abs/pii/S0370269398004547?via\%3Dihub}{Phys.\ Lett.\ B {\bf 429} (1998) 369}     
  
\bibitem{Cacciari:2008gp}
  M.~Cacciari, G.~P.~Salam and G.~Soyez,
  \href{https://dx.doi.org/10.1088/1126-6708/2008/04/063}{JHEP {\bf 0804} (2008) 063}     
    
\bibitem{Campbell:2004ch}
  J.~M.~Campbell, R.~K.~Ellis and F.~Tramontano,
  \href{https://dx.doi.org/10.1103/PhysRevD.70.094012}{Phys.\ Rev.\ D {\bf 70} (2004) 094012}     

\bibitem{Bevilacqua:2015qha}
  G.~Bevilacqua, H.~B.~Hartanto, M.~Kraus and M.~Worek,
  \href{https://journals.aps.org/prl/abstract/10.1103/PhysRevLett.116.052003}{Phys.\ Rev.\ Lett.\  {\bf 116} (2016) no.5,  052003}    

\bibitem{Bevilacqua:2016jfk}
  G.~Bevilacqua, H.~B.~Hartanto, M.~Kraus and M.~Worek,
  \href{https://dx.doi.org/10.1007/JHEP11\%282016\%29098}{JHEP {\bf 1611} (2016) 098}     

\bibitem{Bevilacqua:2018dny}
  G.~Bevilacqua, H.~B.~Hartanto, M.~Kraus, T.~Weber and M.~Worek,
  \href{https://link.springer.com/article/10.1007\%2FJHEP01\%282019\%29188}{JHEP {\bf 1901} (2019) 188}    

\bibitem{Bevilacqua:2019cvp}
  G.~Bevilacqua, H.~B.~Hartanto, M.~Kraus, T.~Weber and M.~Worek,
  \href{https://link.springer.com/article/10.1007\%2FJHEP11\%282019\%29001}{JHEP {\bf 1911} (2019) 001}    
  
\bibitem{Fadin:1993kt}
  V.~S.~Fadin, V.~A.~Khoze and A.~D.~Martin,
  \href{https://www.sciencedirect.com/science/article/pii/0370269394908370?via\%3Dihub}{Phys.\ Lett.\ B {\bf 320} (1994) 141}   
  
  \bibitem{Kauer:2001sp}
  N.~Kauer and D.~Zeppenfeld,
  \href{https://journals.aps.org/prd/abstract/10.1103/PhysRevD.65.014021}{Phys.\ Rev.\ D {\bf 65} (2002) 014021}     
  
\bibitem{Denner:2012mx}
  A.~Denner, S.~Dittmaier, S.~Kallweit and S.~Pozzorini,
  \href{https://pos.sissa.it/151/015}{PoS LL {\bf 2012} (2012) 015}    

\end{thebibliography}
\end{document}